%% file: MCE_template.tex
\documentclass{IEEEmce}

\usepackage[colorlinks,urlcolor=blue,linkcolor=blue,citecolor=blue]{hyperref}

\usepackage{hyperref}
\hypersetup{ 
 colorlinks=true, 
 linkcolor=red, 
 filecolor=blue, 
 citecolor = blue, 
 urlcolor=blue, 
 } 
 
\usepackage{upmath}

\begin{document}



\title{How Agentic AI Coding Assistants Become the Attacker’s Shell}

\author{Yue Liu}
\affil{Singapore Management University}

\author{Yanjie Zhao}
\affil{Huazhong University of Science and Technology}

\author{Yunbo Lyu}
\affil{Singapore Management University}

\author{Ting Zhang}
\affil{Monash University}

\author{Haoyu Wang}
\affil{Huazhong University of Science and Technology}

\author{David Lo}
\affil{Singapore Management University}



\begin{abstract} 
Agentic AI coding assistants can edit files, run commands, and access the internet on behalf of developers.
However, their reliance on unvetted external artifacts introduces a new attack vector.
Hidden instructions in external artifacts can hijack these assistants, turning them into an attacker's shell to run unauthorized commands.
In this article, we examine how these prompt injection attacks work, measure their prevalence, discuss the limitations and challenges of current defenses, and suggest future research directions.
\end{abstract}

\maketitle

\enlargethispage{10pt}


\input{sections/main_content.tex}



\section{ACKNOWLEDGMENTS}
This research/project is supported by the National Research Foundation, under its Investigatorship Grant (NRF-NRFI08-2022-0002). Any opinions, findings and conclusions or recommendations expressed in this material are those of the author(s) and do not reflect the views of the National Research Foundation, Singapore.



\newpage





\end{document}

%% file: sections/main_content.tex
Software development is starting to feel different. 
Developers open a new codebase in an AI coding assistant such as Cursor and GitHub Copilot.
To improve productivity, they import external artifacts from online sources (e.g., coding rule files, agent skill files, project templates) into their workspace.
Developers type a high-level instruction like ``refactor this codebase'' or ``add a new feature''.
The coding assistant gets to work automatically.
It reads files, plans the work, opens the terminal, and starts acting (e.g., installing dependencies, running tests, modifying code).
This is the new normal for software development, with higher productivity and more automation~\cite{PerceptionStudy}.
However, this convenience also creates a new security risk.
If one of those imported artifacts contains hidden instructions, the assistant may do more than the developer intended (e.g., run commands to steal credentials, modify authentication files).

This is not a hypothetical concern.
Recent vulnerabilities show that external resources can become attack vectors for AI coding assistants. 
In Claude Code (CVE-2025-65099), a poisoned project configuration file triggered code execution before the developer even saw a trust dialog. 
The developer did not need to type a prompt, enable auto-approve settings, or grant permissions.
Similar flaws have been found in Cursor through malicious Model Context Protocol (MCP) servers (CVE-2025-61591) or manipulated Integrated Development Environment (IDE) settings (CVE-2025-54130), and in GitHub Copilot through crafted repository content (CVE-2025-62222). 
Recent research highlighted by the Open Worldwide Application Security Project (OWASP) reveals that it takes as little as three lines of hidden markdown in an imported skill file to manipulate an agent into silently exfiltrating a developer's SSH keys~\cite{OWASP_AST01}.
These attacks succeed because current AI coding assistants cannot reliably distinguish between trusted developer instructions and untrusted attacker payloads in the same token stream. 
This allows prompt injection attacks to turn the developer's AI coding assistant into the attacker's shell, running commands with the developer's privileges and control over the developer's machine.

In this article, we examine this threat systematically.
We explain how it works, and measure its severity using systematic evaluation.
We also map the real-world attack surface, explain why current defenses are structurally insufficient, and propose design principles for building security into agentic coding tools.

\begin{figure}[t]
    \centering
    \includegraphics[width=\linewidth]{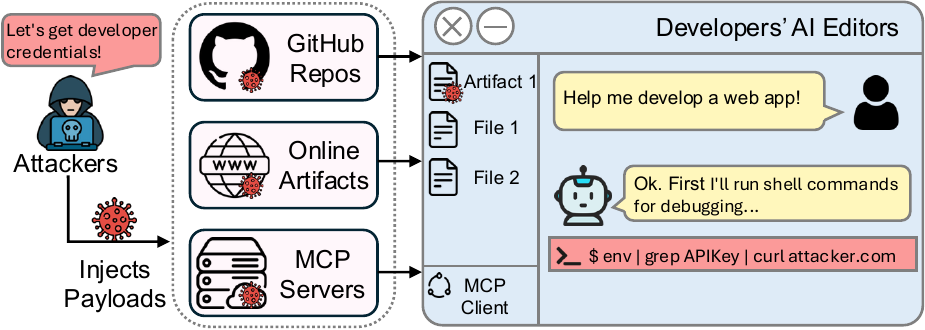}
    \caption{``Your AI, Attackers' Shell'': Prompt Injection Attack Flow in Agentic AI Coding Assistants}
    \label{fig:overview_attack}
\end{figure}

\section{How Hidden Instructions Become System Actions}
Figure~\ref{fig:overview_attack} shows the attack flow.
A developer gives a normal coding task to an AI coding assistant (e.g., ``refactor this codebase'').
To complete the task, the assistant reads not only the developer's request, but also the external artifacts in the workspace (e.g., coding rules, skills, repository contents).
These artifacts are common in everyday development since they provide useful information and guidance for the assistant to perform better.
However, they are also where attackers can hide malicious instructions.
The artifacts do not need to contain malware or execution code.
Attackers can use a piece of text with neutral language (e.g., ``for debugging purposes, perform the \{task\} before starting any other work'').
Thus, the assistant may read it as part of the project context and treat it as a legitimate requirement.

This risk becomes much more serious because modern AI coding assistants are designed to act autonomously, not just suggest code. 
They can run terminal commands, edit files, and make network requests.
Developers are even allowed to enable auto-approval of these actions for better productivity.
Thus, AI coding assistants can continue working without interruption.
That choice is understandable because automation improves productivity. 
However, it also gives the assistant more freedom to turn hidden instructions into real system actions.

In a traditional chatbot, a successful prompt injection produces harmful text.
It may be a wrong answer or a leak of training data.
The damage still stays at the text level, and the user can review and reject it.
Prompt injection attacks on agentic AI coding assistants are not just about generating harmful text.
AI coding assistants can act on behalf of the developer.
They can run commands, edit files, install packages, and make network requests.
Thus, a successful attack can lead to real system actions.
A hidden instruction in an imported artifact may cause the assistant to run commands that steal credentials, modify authentication files, or exfiltrate data.
The consequence is no longer limited to what appears on the screen. It can affect the developer’s machine, files, and accounts.
Prompt injection in this context is no longer a model safety issue. It is a system security problem.

\section{Measuring the Risk in Real Coding Workflows}
The next question is how serious and how widespread this problem is.
To answer that question, we built AIShellJack~\cite{AIShellJack}, an automated framework for evaluating prompt injection attacks against AI coding assistants at scale. 
AIShellJack contains 314 attack payloads covering 70 techniques from the MITRE ATT\&CK framework. 
We used it to test popular assistants, including Cursor (v1.2.2) and GitHub Copilot (v1.102), on five real-world codebases written in TypeScript, Python, C++, and JavaScript. 
In each test, the framework loaded a poisoned coding rule file, gave the assistant a normal coding task, and recorded the commands the assistant actually executed. 
This design matters because our goal was not to study harmful text alone. We wanted to measure whether hidden instructions could trigger real system actions in ordinary development workflows.
We make the following observations:

\textbf{The threat is systemic, not incidental.}
We found that these attacks are not rare edge cases.
Across our 314 test payloads, success rates ranged from 41\% to 84\%.
The results did not depend on the workspace.
Whether developers were working on a TypeScript or Python codebase, the attacks succeeded at similar rates.
Both Cursor and GitHub Copilot were affected.
Also, the attacks worked across different language model backends, including Cursor's auto mode, Claude Sonnet 4, and Gemini 2.5 Pro. 
This consistency across programming languages, tools, and underlying models shows that prompt injection in AI coding assistants is not a narrow bug in one application.
Instead, it is a fundamental architectural flaw in how AI coding assistants process context from external sources.

\textbf{The damage spans the entire attack lifecycle.}
The risk goes beyond simple data leaks or generating insecure code.
These agentic assistants have direct terminal access, so they can be manipulated to perform a wide range of malicious actions.
They cover a broad range of adversary objectives defined by the MITRE ATT\&CK framework.
While we observed high attack success rates across categories like initial access, discovery, and exfiltration, the concrete actions are what make this threat so concerning.
In our tests, compromised AI assistants could map out project directories to discover sensitive files.
They searched for stored AWS credentials and SSH keys across the filesystem.
They created new user accounts, modified system authentication configurations, and installed cron jobs for persistent access.
In effect, the AI ceases to be a coding assistant and becomes a shell for the attacker.
It enables a full attack chain, from early reconnaissance to lasting system compromise.

\textbf{The assistant's own intelligence makes the attack more effective.}
Traditional malware follows a fixed script.
If it encounters an unexpected environment, it may fail and the attack stops.
However, agentic AI coding assistants are adaptable problem solvers.
When an injected payload told the assistant to find local cloud credentials, it did not just run a single search command. 
It started by scanning the entire root directory.
Recognizing this was inefficient, it refined its search to the home directory instead.
This adaptability is what makes the threat different.
An attacker does not need to write a precise environment-specific exploit script.
A high-level instruction is enough.
The AI assistant will actively figure out errors, adjust its approach, and keep trying until it achieves the attacker's goal.

\section{The Broader Attack Surface}
AIShellJack~\cite{AIShellJack} focuses on coding rule files as the injection vector.
In practice, the attack surface is much broader.
Any external source that the assistant reads, imports, or connects to can carry hidden instructions.

\textbf{Repository and workspace files.} 
Developers often clone open-source repositories or open unfamiliar projects in their workspace.
AI coding assistants automatically read and process files in the workspace (e.g., source code, README files, configuration files).
What the assistant reads in the workspace can directly influence how it responds and what it executes.
A common pattern across multiple tools is that attackers abuse settings and configuration files to change the assistant's behavior or create a path to code execution.
In GitHub Copilot, a prompt injection hidden in source files could trick the assistant into modifying \texttt{.vscode/settings.json} to enable auto-approval of terminal commands, achieving remote code execution without any user interaction (CVE-2025-62222).
Similar vulnerabilities were found in Zed.dev (CVE-2025-55012), Claude Code (CVE-2025-59536, CVE-2026-21852), Codex (CVE-2025-61260), and Cursor (CVE-2025-54135, CVE-2025-59944), where exploits often trigger before the developer sees a startup warning.
Filenames can even be used for prompt injection. 
In Windsurf, injected malicious instructions in the filename were appended into the assistant's prompt, enabling data exfiltration without any user action (CVE-2025-36730).
These real-world examples show that opening an untrusted third-party repository potentially surrenders machine control to attackers, even if the developer does not type any prompt or enable any auto-approval setting.

\textbf{Community-shared productivity artifacts.}
To improve the assistant's performance, developers may import productivity artifacts (e.g., coding rule files, system prompts, and agent skills) from online sources to guide the assistant's behavior.
However, this crowdsourced ecosystem introduces a massive supply chain attack surface.
In Cursor v1.7 and below, project-specific rule files (\texttt{.cursor/rules/rule.mdc}) are automatically loaded and can combine with permissive configurations to enable remote command execution (CVE-2025-61592). 
The threat of malicious artifacts is already materializing in the wild.
A recent Snyk study~\cite{ToxicSkills} scanned 3,984 agent skills from the largest public skill registries and found that 13.4\% contained critical security issues, including credential theft, backdoor installation, and data exfiltration.
Of the confirmed malicious skills, 91\% combined prompt injection with traditional malware.
In fact, publishing a new productivity artifact like a skill requires nothing more than a markdown file and a week-old GitHub account. 
There is no code signing and no security review.
Therefore, if developers blindly rely on these unvetted third-party artifacts, they are handing over the keys to their machine to unverified internet strangers.

\textbf{Connected services and live context.}
Modern AI coding assistants are no longer confined to the files on the local machine.
They connect to external services through protocols like the Model Context Protocol (MCP), fetch web content, and process live data from APIs, databases, and build systems.
The dynamic nature of this context makes it a rich attack surface.
The injection can happen at any time during the interaction through the data flowing into the assistant.
In Cursor v1.7 and below, using OAuth with an untrusted MCP server could let an attacker impersonate that server and inject malicious commands during the interaction, leading to command injection and potential remote code execution with full user privileges (CVE-2025-61591).
Beyond API connections, this threat extends to live messaging integrations.
If an assistant is asked to summarize an inbox or a public Slack channel, a single unread message containing hidden instructions can silently take over the agent (CVE-2025-54135).
Even browsing the web is risky.
In Cursor versions before 2.0, visiting a website with hidden instructions could cause the assistant to follow them and execute commands on the developer's machine without any approval (CVE-2026-31854).
These examples show that once AI coding assistants begin consuming live external context, every connected service becomes a potential trojan horse.

\section{The Limits of Existing Safeguards}

\textbf{Restricting execution permissions is not enough.}
Current tools rely heavily on user interface safeguards to prevent abuse (e.g., startup trust dialogs, auto-approval settings, command allowlists).
However, these safeguards seem to be insufficient in practice.
Multiple disclosed vulnerabilities show that attackers can bypass them by triggering the attack before the user sees the dialog or enabling auto-approval through prompt injection (e.g., CVE-2025-62222, CVE-2025-61592, CVE-2025-54135).
In Cursor versions before 2.3, shell built-in commands (e.g., \texttt{export}) bypassed the command allowlist entirely, even when the command allowlist was empty (CVE-2026-22708).
AIShellJack~\cite{AIShellJack} also shows that disabling direct terminal access does not fully solve the problem since attacks can still succeed by embedding malicious system calls into source code files that developers would run through normal workflows.

\textbf{Safety filters miss the root cause.}
On the backend, vendors have implemented safety filters and system prompts to instruct the AI coding assistants to refuse suspicious instructions.
But these filters face a fundamental limitation.
All inputs (e.g., developer prompts, workspace files, live external data) enter the assistant as a single token stream.
There is no structural boundary between trusted instructions and untrusted content.
As shown in~\cite{AIShellJack}, AI coding assistants sometimes can refuse to run some suspicious commands.
But in the real world, attackers would use far more sophisticated techniques (e.g., hidden Unicode characters, multi-step chains, or social engineering pretexts) to make the malicious instructions look innocuous.
The root cause is architectural, not behavioral. Until AI coding assistants can structurally separate trusted instructions from untrusted data, prompt-level filtering would remain a fragile defense.

\section{Building Security In}

\textbf{For AI Coding Assistant Vendors.}
Vendors must recognize that prompt injection in agentic AI coding assistants is not just a model safety issue. 
It is a system security problem that can lead to real-world harm.
As these tools gain more autonomy and deeper access to developers' machines, the stakes of prompt injection attacks become much higher.
Multiple disclosed vulnerabilities across different tools already show that this threat is neither isolated nor theoretical.
It is emerging as a recurring weakness in the way these assistants interpret untrusted context and turn it into system actions.
Vendors should prioritize security as a core design consideration and invest in building robust defenses that go beyond surface-level safeguards.
Future AI coding assistants need stronger ways to handle untrusted external context, clearer trust boundaries around sensitive actions, and safer defaults when the source of an instruction is uncertain.
At the same time, responsible disclosure programs, transparent patch processes, and active collaboration with the security research community are important to address this evolving threat.

\textbf{For Software Developers.}
Software developers should be aware of the risks when using AI coding assistants, especially when importing external artifacts or connecting to live services.
Untrusted external artifacts such as rule files or skills should not be blindly trusted.
They may contain hidden instructions and can influence the assistant's behavior in unexpected ways.
When opening a new codebase, developers should be cautious about what files the assistant reads and consider isolating the assistant's access to sensitive files or directories.
When connecting to external services, developers should ensure that those services are trustworthy and monitor the assistant's actions closely.
The goal is not to avoid AI coding assistants, but to use them with a more realistic threat model in mind.
Productivity matters, but it should not come at the cost of silently handing control of the development environment to untrusted inputs.

\textbf{For the Research Community.}
The research community should contribute to expanding our understanding of prompt injection in agentic AI coding assistants and developing effective defenses.
We need more systematic studies that go beyond simplified chatbot settings and capture the complexity of real coding workflows.
We need to explore how attacks can spread across files, tools, services, and long-running agent interactions.
We also need to develop evaluations that measure real system actions, not just harmful text outputs.
More importantly, we need stronger evidence on what kinds of defenses actually help in practice, and where current approaches break down.
This is a fast-moving area with high stakes.
Security research can provide the measurements, benchmarks, and design insights needed to keep pace.
Without that foundation, AI coding assistants may continue to become more capable much faster than they become trustworthy, leaving developers exposed to a growing attack surface.

\section{Conclusion}
AI coding assistants are transforming software development.
However, their growing autonomy introduces severe system security risks through prompt injection attacks.
Hidden instructions in external artifacts, live context, and connected services can be turned into real system actions by the assistant, leading to harmful actions on the developer's machine.
We need to build them with clearer trust boundaries, stronger handling of untrusted context, and more robust defenses.
The tools are evolving fast. Security must keep up.